# Ideas by Statistical Mechanics (ISM)[†]

*Lester Ingber*
**Lester Ingber Research (LIR)**
<ingber@ingber.com> <ingber@alumni.caltech.edu> [http://www.ingber.com]

**Abstract**

Ideas by Statistical Mechanics (ISM) is a generic program to model evolution and propagation of ideas/patterns throughout populations subjected to endogenous and exogenous interactions. The program is based on the author's work in Statistical Mechanics of Neocortical Interactions (SMNI), and uses the author's Adaptive Simulated Annealing (ASA) code for optimizations of training sets, as well as for importance-sampling to apply the author's copula financial risk-management codes, Trading in Risk Dimensions (TRD), for assessments of risk and uncertainty. This product can be used for decision support for projects ranging from diplomatic, information, military, and economic (DIME) factors of propagation/evolution of ideas, to commercial sales, trading indicators across sectors of financial markets, advertising and political campaigns, etc.

It seems appropriate to base an approach for propagation of ideas on the only system so far demonstrated to develop and nurture ideas, i.e., the neocortical brain. A statistical mechanical model of neocortical interactions, developed by the author and tested successfully in describing short-term memory and EEG indicators, is the proposed model. ISM develops subsets of macrocolumnar activity of multivariate stochastic descriptions of defined populations, with macrocolumns defined by their local parameters within specific regions and with parameterized endogenous inter-regional and exogenous external connectivities. Parameters with a given subset of macrocolumns will be fit using ASA to patterns representing ideas. Parameters of external and inter-regional interactions will be determined that promote or inhibit the spread of these ideas. Tools of financial risk management, developed by the author to process correlated multivariate systems with differing non-Gaussian distributions using modern copula analysis, importance-sampled using ASA, will enable bona fide correlations and uncertainties of success and failure to be calculated. Marginal distributions will be evolved to determine their expected duration and stability using algorithms developed by the author, i.e., PATHTREE and PATHINT codes.

# Contents

# 1. Significance of Problem

A briefing [2] demonstrates the breadth and depth complexity required to address real diplomatic, information, military, economic (DIME) factors for the propagation/evolution of ideas through defined populations. An open mind would conclude that it is possible that multiple approaches may be required for multiple decision makers in multiple scenarios. However, it is in the interests of multiple decision-makers to as much as possible rely on the same generic model for actual computations. Many users would have to trust that the coded model is faithful to process their input.

Ideas by Statistical Mechanics (ISM) can be developed to address these issues. [ism (noun): A belief (or system of beliefs) accepted as authoritative by some group or school. A doctrine or theory; especially, a wild or visionary theory. A distinctive doctrine, theory, system, or practice.]

## 1.1. Bottom-Up versus Top-Down

The concept of "memes" is an example of an approach to deal with DIME factors [68].

The meme approach, using a reductionist philosophy of evolution among genes, is reasonably contrasted to approaches emphasizing the need to include relatively global influences of evolution [69].

It seems appropriate to base an approach for propagation of ideas on the only system so far demonstrated to develop and nurture ideas, i.e., the neocortical brain. In the present context, the author's approach, using guidance from his statistical mechanics of human neocortical interactions (SMNI), developed in a series of about 30 published papers from 1981-2001 [13-15,19,28,32,34,40,42,43,45], also addresses long-standing issues of information measured by electroencephalography (EEG) as arising from bottom-up local interactions clusters of thousands to tens of thousands of neurons interacting via short-ranged fibers), or top-down influences of global interactions (mediated by long-ranged myelinated fibers). SMNI does this by including both local and global interactions as being necessary to develop neocortical circuitry.

## 1.2. Cost Functions for Ideas

Computational approaches developed to process different approaches to modeling phenomena must not be confused with the models of these phenomena. For example, the meme approach lends it self well to a computational scheme in the spirit of genetic algorithms (GA). The cost/objective function that describes the phenomena of course could be processed by any other sampling technique such as simulated annealing (SA). One comparison [57] demonstrated the superiority of SA over GA on cost/objective functions used in a GA database. That study used Very Fast Simulated Annealing (VFSR), created by the author for military simulation studies [25], which has evolved into Adaptive Simulated Annealing (ASA) [29]. However, it is the author's experience that the Art and Science of sampling complex systems requires tuning expertise of the researcher as well as good codes, and GA or SA likely would do as well on cost functions for this study.

A very important issue is for this projects is to develop cost functions for this study, not only how to fit or process them.

## 1.3. Inclusion of non-Gaussian Correlated Systems

This proposal includes application of methods of portfolio risk analysis to such statistical systems. There are often two kinds of errors committed in multivariate risk analyses: (E1) Although the distributions of variables being considered are not Gaussian (or not tested to see how close they are to Gaussian), standard statistical calculations appropriate *only* to Gaussian distributions are employed. (E2) Either correlations among the variables are ignored, or the mistakes committed in (E1) — incorrectly assuming variables are Gaussian — are compounded by calculating correlations as if all variables were Gaussian.

The harm in committing errors E1 and E2 can be fatal — fatal to the analysis and/or fatal to people acting in good faith on the basis of these risk assessments. Risk is measured by tails of distributions. So, if the tails of some variables are much fatter or thinner than Gaussian, the risk in committing E1 can be quite terrible. Many times systems are pushed to and past desired levels of risk when several variables become highly correlated, leading to extreme dependence of the full system on the sensitivity of these variables. It is very important not to commit E2 errors.

The Trading in Risk Dimensions (TRD) project addresses these issues in the context of financial risk management, but the tools and codes are generic.

### 1.4. Other Alternatives

There are multiple other alternative works being conducted world-wide that must be at least kept in mind while developing and testing models of evolution/propagation of ideas in defined populations: A study on a (too) simple algebraic model of opinion formation concluded that the only final opinions are extremal ones [1]. A study of the influence on chaos on opinion formation, using a simple algebraic model, concluded that contrarian opinion could persist and be crucial in close elections, albeit the authors were careful to note that most real populations probably do not support chaos [5]. A limited review of work in social networks illustrates that there are about as many phenomena to be explored as there are disciplines ready to apply their network models. [67]

## 2. Technical Objectives

### 2.1. Architecture for Selected Model

The primary objective is to deliver a computer model that contains the following features: (1) A multivariable space will be defined to accommodate populations. (2) A cost function over the population variables in (1) will be defined to explicitly define a pattern that can be identified as an Idea. (3) As subset of the population will be used to fit parameters — e.g, coefficients of variables, connectivities to patterns, etc. — to an Idea, using the cost function in (2). (4) Connectivity of the population in (3) will be made to the rest of the population. Investigations will be made to determine what endogenous connectivity is required to stop or promote the propagation of the Idea into other regions of the population. (5) External forces, e.g., acting only on specific regions of the population, will be introduced, to determine how these exogenous forces may stop or promote the propagation of an Idea.

## 3. Work Plan

### 3.1. Application of SMNI Model

A statistical mechanical model of neocortical interactions, developed by the author and tested successfully in describing short-term memory (STM) and in training and testing EEG indicators, is the model proposed to be used here to address DIME factors for the propagation/evolution of Ideas through defined populations.

The approach here is to develop subsets of Ideas/macrocolumnar activity of multivariate stochastic descriptions of defined populations (of a reasonable but small population sample, e.g., of 100–1000), with macrocolumns defined by their local parameters within specific regions (larger samples of populations) and with parameterized long-ranged inter-regional and external connectivities. Parameters of a given subset of macrocolumns will be fit using ASA to patterns representing Ideas, akin to acquiring hard-wired long-term memory (LTM) patterns. Parameters of external and inter-regional interactions will be determined that promote or inhibit the spread of these Ideas, by determining the degree of fits and overlaps of probability distributions relative to the seeded macrocolumns.

That is, the same Ideas/patterns may be represented in other than the seeded macrocolumns by local confluence of macrocolumnar and long-ranged firings, akin to STM, or by different hard-wired parameter LTM sets that can support the same local firings in other regions (possible in nonlinear systems). SMNI also calculates how STM can be dynamically encoded into LTM [14].

Small populations in regions will be sampled to determine if the propagated Idea(s) exists in its pattern space where it did exist prior to its interactions with the seeded population. SMNI derives nonlinear functions as arguments of probability distributions, leading to multiple STM, e.g., $7 \pm 2$ for auditory memory capacity. Some investigation will be made into nonlinear functional forms other than those derived for SMNI, e.g., to have capacities of tens or hundreds of patterns, etc.

## 3.2. Application of TRD Analysis

Tools of financial risk management, developed by the author to process correlated multivariate systems with differing non-Gaussian distributions using modern copula analysis, importance-sampled using ASA, will enable bona fide correlations and uncertainties of success and failure to be calculated [51]. Marginal distributions will be evolved to determine their expected duration and stability using algorithms developed by the author, i.e., PATHTREE [52] and PATHINT [44] codes.

## 4. Related Work

### 4.1. Statistical Mechanics of Neocortical Interactions (SMNI)

#### 4.1.1. Application to Proposed Project

Neocortex has evolved to use minicolumns of neurons interacting via short-ranged interactions in macrocolumns, and interacting via long-ranged interactions across regions of macrocolumns. This common architecture processes patterns of information within and among different regions of sensory, motor, associative cortex, etc. Therefore, the premise of this proposal is that this is a good model to describe and analyze evolution/propagation of Ideas among defined populations.

Relevant to this study is that a spatial-temporal lattice-field short-time conditional multiplicative-noise (nonlinear in drifts and diffusions) multivariate Gaussian-Markovian probability distribution is developed faithful to neocortical function/physiology. Such probability distributions are a basic input into the approach used here. The SMNI model was the first physical application of a nonlinear multivariate calculus developed by other mathematical physicists in the late 1970's to define a statistical mechanics of multivariate nonlinear nonequilibrium systems [9,62].

#### 4.1.2. SMNI Tests on STM and EEG

The author has developed a statistical mechanics of neocortical interactions (SMNI) for human neocortex, building from synaptic interactions to minicolumnar, macrocolumnar, and regional interactions in neocortex. Since 1981, a series of papers on the statistical mechanics of neocortical interactions (SMNI) has been developed to model columns and regions of neocortex, spanning mm to cm of tissue. Most of these papers have dealt explicitly with calculating properties of STM and scalp EEG in order to test the basic formulation of this approach [12-15,18,19,21,27,28,32,34,35,37,38,40,42,55,56].

The SMNI modeling of local mesocolumnar interactions (convergence and divergence between minicolumnar and macrocolumnar interactions) was tested on STM phenomena. The SMNI modeling of macrocolumnar interactions across regions was tested on EEG phenomena.

#### 4.1.3. SMNI Description of STM

SMNI studies have detailed that maximal numbers of attractors lie within the physical firing space of $M^G$, where $G$ = {Excitatory, Inhibitory} minicolumnar firings, consistent with experimentally observed capacities of auditory and visual STM, when a "centering" mechanism is enforced by shifting background noise in synaptic interactions, consistent with experimental observations under conditions of selective attention [15,19,32,56,65]. This leads to all attractors of the short-time distribution lying along a diagonal line in $M^G$ space, effectively defining a narrow parabolic trough containing these most likely firing states. This essentially collapses the 2 dimensional $M^G$ space down to a one-dimensional space of most importance. Thus, the predominant physics of STM and of (short-fiber contribution to) EEG phenomena takes place in a narrow "parabolic trough" in $M^G$ space, roughly along a diagonal line [15].

These calculations were further supported by high-resolution evolution of the short-time conditional-probability propagator using PATHINT [56]. SMNI correctly calculated the stability and duration of STM, the primacy versus recency rule, random access to memories within tenths of a second as observed, and the observed $7 \pm 2$ capacity rule of auditory memory and the observed $4 \pm 2$ capacity rule of visual memory.

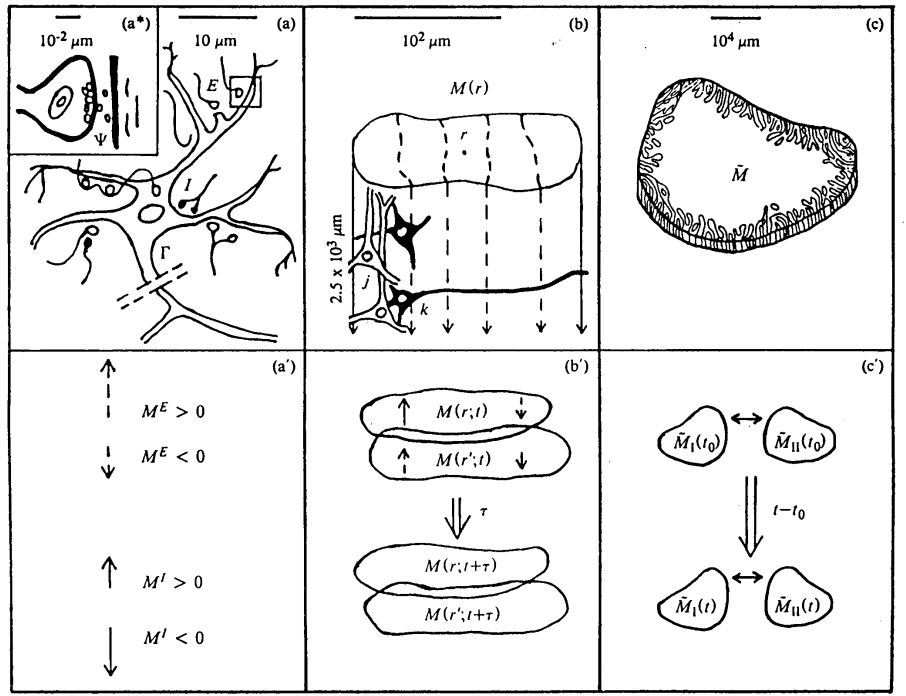

Fig. 1. Illustrated are three biophysical scales of neocortical interactions: (a)-($a^*$)-(a') microscopic neurons; (b)-(b') mesocolumnar domains; (c)-(c') macroscopic regions. SMNI has developed appropriate conditional probability distributions at each level, aggregating up from the smallest levels of interactions. In ($a^*$) synaptic inter-neuronal interactions, averaged over by mesocolumns, are phenomenologically described by the mean and variance of a distribution $\Psi$. Similarly, in (a) intraneuronal transmissions are phenomenologically described by the mean and variance of $\Gamma$. Mesocolumnar averaged excitatory ($E$) and inhibitory ($I$) neuronal firings are represented in (a'). In (b) the vertical organization of minicolumns is sketched together with their horizontal stratification, yielding a physiological entity, the mesocolumn. In (b') the overlap of interacting mesocolumns is sketched. In (c) macroscopic regions of neocortex are depicted as arising from many mesocolumnar domains. (c') sketches how regions may be coupled by long-ranged interactions.

SMNI also calculates how STM patterns may be encoded by dynamic modification of synaptic parameters (within experimentally observed ranges) into long-term memory patterns (LTM) [14].

### 4.1.4. SMNI Description of EEG

Using the power of this formal structure, sets of EEG and evoked potential data from a separate NIH study, collected to investigate genetic predispositions to alcoholism, were fitted to an SMNI model on a lattice of regional electrodes to extract brain "signatures" of STM [40,42]. Each electrode site was represented by an SMNI distribution of independent stochastic macrocolumnar-scaled $M^G$ variables, interconnected by long-ranged circuitry with delays appropriate to long-fiber communication in neocortex. The global optimization algorithm ASA was used to perform maximum likelihood fits of Lagrangians defined by path integrals of multivariate conditional probabilities. Canonical momenta indicators (CMI) were thereby derived for individual's EEG data. The CMI give better signal recognition than the raw data, and were used to advantage as correlates of behavioral states. In-sample data was used for training [40], and out-of-sample data was used for testing [42] these fits.

These results gave strong quantitative support for an accurate intuitive picture, portraying neocortical interactions as having common algebraic physics mechanisms that scale across quite disparate spatial scales and functional or behavioral phenomena, i.e., describing interactions among neurons, columns of neurons, and regional masses of neurons.

### 4.1.5. Generic Mesoscopic Neural Networks

SMNI was applied to propose a parallelized generic mesoscopic neural networks (MNN) [28], adding computational power to a similar paradigm proposed for target recognition [17].

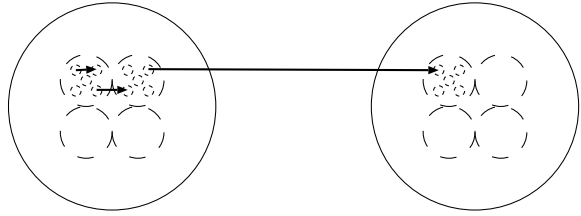

Fig. 2. "Learning" takes place by presenting the MNN with data, and parametrizing the data in terms of the "firings," or multivariate $M^G$ "spins." The "weights," or coefficients of functions of $M^G$ appearing in the drifts and diffusions, are fit to incoming data, considering the joint "effective" Lagrangian (including the logarithm of the prefactor in the probability distribution) as a dynamic cost function. This program of fitting coefficients in Lagrangian uses methods of ASA.

"Prediction" takes advantage of a mathematically equivalent representation of the Lagrangian path-integral algorithm, i.e., a set of coupled Langevin rate-equations. A coarse deterministic estimate to "predict" the evolution can be applied using the most probable path, but PATHINT has been used. PATHINT, even when parallelized, typically can be too slow for "predicting" evolution of these systems. However, PATHTREE is much faster.

The present project uses the same concepts, having sets of multiple variables define macrocolumns with a region, with long-ranged connectivity to other regions. Each macrocolumn has its own parameters, which define sets of possible patterns. Ultimately, ISM of course would not use functional relationships developed solely in neocortex, but rather those more appropriate to a given population.

### 4.1.6. On Chaos in Neocortex

There are many papers on the possibility of chaos in neocortical interactions. While this phenomena may have some merit when dealing with small networks of neurons, e.g., in some circumstances such as epilepsy, these papers generally have considered only too simple models of neocortex.

The author took a model of chaos that might be measured by EEG, developed and published by some colleagues, but adding background stochastic influences and parameters that were agreed to better model neocortical interactions. The resulting multivariate nonlinear conditional probability distribution was propagated many thousands of epochs, using the authors PATHINT code, to see if chaos could exist and persist under such a model [58]. There was absolutely no measurable instance of chaos surviving in this more realistic context.

### 4.1.7. Mathematical Development

Some of the algebra behind SMNI depicts variables and distributions that populate each representative macrocolumn in each region.

A derived mesoscopic Lagrangian $L_M$ defines the short-time probability distribution of firings in a minicolumn, composed of ~$10^2$ neurons, given its just previous interactions with all other neurons in its macrocolumnar surround. $G$ is used to represent excitatory ($E$) and inhibitory ($I$) contributions. $\bar{G}$ designates contributions from both $E$ and $I$.

$$P_M = \prod_G P_M^G[M^G(r; t+\tau)|M^{\bar{G}}(r'; t)]$$

$$= \sum_{\sigma_j} \delta\left(\sum_{jE} \sigma_j - M^E(r; t+\tau)\right) \delta\left(\sum_{jI} \sigma_j - M^I(r; t+\tau)\right) \prod_j^N p_{\sigma_j}$$

$$\approx \prod_G (2\pi\tau g^{GG})^{-1/2} \exp(-N\tau \underline{L}_M^G) \, ,$$

$$P_M \approx (2\pi\tau)^{-1/2} g^{1/2} \exp(-N\tau \underline{L}_M) \, ,$$

$$\underline{L}_M = \underline{L}_M^E + \underline{L}_M^I = (2N)^{-1}(\dot{M}^G - g^G) g_{GG'}(\dot{M}^{G'} - g^{G'}) + M^G J_G/(2N\tau) - \underline{V}' \, ,$$

$$\underline{V}' = \sum_G \underline{V}''^G_{G'} (\rho \nabla M^{G'})^2 \, ,$$

$$g^G = -\tau^{-1}(M^G + N^G \tanh F^G) \, , \quad g^{GG'} = (g_{GG'})^{-1} = \delta_G^{G'} \tau^{-1} N^G \operatorname{sech}^2 F^G \, , \quad g = \det(g_{GG'}) \, ,$$

$$F^G = \frac{(V^G - a^{|G|}_{G'} v^{|G|}_{G'} N^{G'} - \frac{1}{2} A^{|G|}_{G'} v^{|G|}_{G'} M^{G'})}{(\pi[(v^{|G|}_{G'})^2 + (\phi^{|G|}_{G'})^2](a^{|G|}_{G'} N^{G'} + \frac{1}{2} A^{|G|}_{G'} M^{G'}))^{1/2}} \, , \quad a^G_{G'} = \frac{1}{2} A^G_{G'} + B^G_{G'} \, , \tag{1}$$

where $A^G_{G'}$ and $B^G_{G'}$ are minicolumnar-averaged inter-neuronal synaptic efficacies, $v^G_{G'}$ and $\phi^G_{G'}$ are averaged means and variances of contributions to neuronal electric polarizations. $M^{G'}$ and $N^{G'}$ in $F^G$ are afferent macrocolumnar firings, scaled to efferent minicolumnar firings by $N/N* \sim 10^{-3}$, where $N*$ is the number of neurons in a macrocolumn, $\sim 10^5$. Similarly, $A^{G'}_G$ and $B^{G'}_G$ have been scaled by $N*/N \sim 10^3$ to keep $F^G$ invariant. $\underline{V}'$ are mesocolumnar nearest-neighbor interactions.

### 4.1.7.1. Inclusion of Macroscopic Circuitry

The most important features of this development are described by the Lagrangian $\underline{L}^G$ in the negative of the argument of the exponential describing the probability distribution, and the "threshold factor" $F^G$ describing an important sensitivity of the distribution to changes in its variables and parameters.

To more properly include long-ranged fibers, when it is possible to numerically include interactions among macrocolumns, the $J_G$ terms can be dropped, and more realistically replaced by a modified threshold factor $F^G$,

$$F^G = \frac{(V^G - a^{|G|}_{G'} v^{|G|}_{G'} N^{G'} - \frac{1}{2} A^{|G|}_{G'} v^{|G|}_{G'} M^{G'} - a^{\ddagger E}_{E'} v^E_{E'} N^{\ddagger E} - \frac{1}{2} A^{\ddagger E}_{E'} v^E_{E'} M^{\ddagger E})}{(\pi[(v^{|G|}_{G'})^2 + (\phi^{|G|}_{G'})^2](a^{|G|}_{G'} N^{G'} + \frac{1}{2} A^{|G|}_{G'} M^{G'} + a^{\ddagger E}_{E'} N^{\ddagger E} + \frac{1}{2} A^{\ddagger E}_{E'} M^{\ddagger E}))^{1/2}} \, ,$$

$$a^{\ddagger E}_{E'} = \frac{1}{2} A^{\ddagger E}_{E'} + B^{\ddagger E}_{E'} \, . \tag{2}$$

Here, afferent contributions from $N^{\ddagger E}$ long-ranged excitatory fibers, e.g., cortico-cortical neurons, have been added, where $N^{\ddagger E}$ might be on the order of 10% of $N^*$: Of the approximately $10^{10}$ to $10^{11}$ neocortical neurons, estimates of the number of pyramidal cells range from 1/10 to 2/3. Nearly every pyramidal cell has an axon branch that makes a cortico-cortical connection; i.e., the number of cortico-cortical fibers is of the order $10^{10}$.

### 4.1.8. Portfolio of Physiological Indicators (PPI)

The ISM project uses the SMNI distributions as a template for distributions of populations. The TRD project illustrates how such distributions can be developed as a Portfolio of Physiological Indicators (PPI), to calculate risk and uncertainty of functions, e.g., functions of Ideas, dependent on events that impact such populations.

### 4.1.8.1. Multiple Imaging Data

It is clear that the SMNI distributions also can be used to process different imaging data beyond EEG, e.g., also MEG, PET, SPECT, fMRI, etc., where each set of imaging data is used to fit it own set of parameterized SMNI distributions using a common regional circuitry. (Different imaging techniques may have different sensitivities to different synaptic and neuronal activities.) Then, portfolios of these imaging distributions can be developed to describe the total neuronal system, e.g., akin to a portfolio of a basket of markets. For example, this could permit the uncertainties of measurements to be reduced by weighting the contributions of different data sets, etc. Overlaps of distributions corresponding to different subsets of data give numerical specificity to the values of using these subsets.

It is to be expected that better resolution of behavioral events can be determined by joint distributions of different imaging data, rather than by treating each distribution separately.

### 4.1.8.2. Local Versus Global Influences

Another twist on the use of this approach is to better understand the role of local and global contributions to imaging data. EEG data is often collected at different electrode resolutions. Cost functions composed of these different collection-method variables can be used to calculate expectations over their imaging portfolios. For example, relative weights of two scales of collection methods can be fit as parameters, and relative strengths as they contribute to various circuitries can be calculated. This method will be applied to determine the degree of relevance of local and global activity during specific tasks. If some tasks involve circuitry with frontal cortex, then these methods may contribute to the understanding of the role of consciousness.

### 4.1.8.3. Application to ISM

These kinds of applications of SMNI and TRD to PPI have obvious counterparts in ISM. Different kinds of data from populations often lead to different conclusions. A portfolio of distributions from these different data sets permits a better assessment of relative error/uncertainty of these conclusions.

## 4.2. Computational Physics

### 4.2.1. Application to Proposed Project

The author's work in mathematical and computational physics, applying algorithms including those used in this project to applications in several disciplines [20,33,41,52,58], neuroscience [13,15,28,34,38,42], finance [26,44,54,60], general optimization [3,25,30,36,57], and combat analysis [6,31,53,59], illustrate the importance of properly applying these algorithms to the proposed project.

### 4.2.2. Adaptive Simulated Annealing (ASA)

Adaptive Simulated Annealing (ASA) [29] is used to optimize parameters of systems and also to importance-sample variables for risk-management.

ASA is a C-language code developed to statistically find the best global fit of a nonlinear constrained non-convex cost-function over a $D$-dimensional space. This algorithm permits an annealing schedule for "temperature" $T$ decreasing exponentially in annealing-time $k$, $T = T_0 \exp(-ck^{1/D})$. The introduction of re-annealing also permits adaptation to changing sensitivities in the multi-dimensional parameter-space. This annealing schedule is faster than fast Cauchy annealing, where $T = T_0/k$, and much faster than Boltzmann annealing, where $T = T_0/\ln k$. ASA has over 100 OPTIONS to provide robust tuning over many classes of nonlinear stochastic systems.

For example, ASA has ASA_PARALLEL OPTIONS, hooks to use ASA on parallel processors, which were first developed in 1994 when the author of this proposal was Principal Investigator (PI) of National Science Foundation grant DMS940009P, Parallelizing ASA and PATHINT Project (PAPP). Since then these OPTIONS have been used by people in various institutions.

### 4.2.3. PATHINT and PATHTREE

In some cases, it is desirable to develop a time evolution of a short-time conditional probability, e.g., of marginal distributions in this study. Two useful algorithms have been developed and published by the author.

PATHINT [32] motivated the development of PATHTREE [52], an algorithm that permits extremely fast accurate computation of probability distributions of a large class of general nonlinear diffusion processes.

The natural metric of the space is used to first lay down the mesh. The evolving local short-time distributions on this mesh are then dynamically calculated. The short-time probability density gives the correct result up to order $O(\Delta t)$ for any final point $S'$, the order required to recover the corresponding partial differential equation.

PATHINT and PATHTREE have demonstrated their utility in statistical mechanical studies in finance, neuroscience, combat analyses, neuroscience, and other selected nonlinear multivariate systems [44,53,56]. PATHTREE has been used extensively to price financial options [52].

### 4.3. Statistical Mechanics of Combat (SMC)

#### 4.3.1. Application to Proposed Project

The author has experience in several disciplines developing projects requiring developing and fitting nonlinear stochastic algorithms to data, including projects that require developing algorithms for accurate description and analysis of human activity.

#### 4.3.2. Janus Project

During 1988-1989, after a year of preparatory work, as a Professor of Physics with the US Navy, and working with the US Army, the author of this proposal was PI of US Army Contract RLF6L, funded by the Deputy Under Secretary of the Army for Operations Research (DUSA-OR). He led a team of Officers and contractors to successfully baseline Janus(T) — a battalion-level war game with statistical details of performance characteristics of weapons, movement of men and machines across various terrains — to National Training Center (NTC) data obtained in the field [6,22-24].

The Janus project developed fits of data to probability distributions, separately for the data collected at NTC and for the data collected from Janus(T) war games (after the terrain and tactics used at NTC were put into Janus). A Statistical Mechanics of Combat (SMC) was developed, essentially a nonlinear stochastic extension of Lanchester theory of combat, to define a common cost function. The fits were performed using an early variant of ASA, Very Fast Simulated Re-annealing (VFSR) also created by the author [25]. These distributions were evolved in time, to test their sensitivity to the initial fits. A match between the means and variances of the two evolving distributions gave the US Army confidence to use Janus(T) in acquisition and tactics training.

#### 4.3.3. Portfolio of Combat Indicators (PCI)

Many times (multiple runs of) simulation studies are performed to study the influence of a particular technology or set of tactics using varied technologies, within the context of a full scenario of multiple technologies and tactics/strategies.

The PPI project illustrates how multiple distributions, derived from independent fits of such simulations can be developed as a Portfolio Combat Indicators (PCI), to calculate risk and uncertainty of functions of these technologies and/or tactics.

#### 4.3.3.1. Application to ISM

Similar to the utility of PPI to help ground the concept of ISM by a reasonable analogy to phenomena more familiar than ISM, the kinds of applications of PCI have obvious counterparts in ISM. Different kinds of data from populations often lead to different conclusions. A portfolio of distributions from these different data sets permits a better assessment of relative error/uncertainty of these conclusions.

### 4.4. Trading in Risk Dimensions (TRD)

### 4.4.1. Statistical mechanics of Financial Markets (SMFM)

A full real-time risk-managed trading system has been coded by the author using state of the art risk management algorithms, Trading in Risk Dimensions (TRD) [51].

TRD is based largely on previous work in several disciplines, using a formulation similar to that used by the author to develop a multivariate nonlinear nonequilibrium Statistical Mechanics of Financial Markets (SMFM) [16,26,39,47-49]. using powerful numerical algorithms to fit models to data [46]. A published report closest to this project was formulated for a portfolio of options [50].

### 4.4.2. Application to Proposed Project

In the context of this proposal, the concepts of "portfolio" are considered to be extended to the total ensemble of of multiple regions of populations, each having sets of multiple variables. That is, although the each region will have the same kinds of multiple variables, to create a generic system for the project, such variables in different regions will be part of the full set of multivariate nonlinear stochastic variables across all regions. Once the full "portfolio" distribution is developed, various measures of cost or performance can be calculated, in addition to calculating various measure of risk.

The concepts of trading-rule parameters are considered to be extended to parameters that might be included in this work, e.g., to permit some top-level control of weights given to different members of ensembles, or parameters in models that affect their interactions.

It is clear that stochastic financial markets represent a social system of many people willing to risk their money on their beliefs and ideas and on their assumptions of beliefs and ideas of other traders. The concepts of trading rules and portfolio risk-management seem useful to introduce into ISM, beyond tools to determine risk and uncertainty.

#### 4.4.2.1. Standard Code For All Platforms

The ASA and TRD codes are in vanilla C, able to run across all Unix platforms, including Linux and Cygwin under Windows [http://cygwin.com]. Standard Unix scripts are used to facilitate file and data manipulations. For example, output analysis plots — e.g., 20 sub-plots per page, are prepared in batch using RDB (a Perl relational database tool from ftp://ftp.rand.org/RDB-hobbs/), Gnuplot (from http://gnuplot.sourceforge.net/), and other Unix scripts developed by the author.

The judicious use of pre-processing and post-processing of variables, in addition to processing by optimization and importance-sampling algorithms, presents important features to the proposed project beyond simple maximum likelihood estimates based on (quasi-)linear methods of regression usually applied to such systems.

TRD includes design and code required to interface to actual data feeds and execution platforms. Similar requirements might be essential for future use of these approaches in the project proposed here.

As with most complex projects, care must be given to sundry problems that arise. Similar and new such problems can be expected to arise in this project as well.

#### 4.4.2.2. Gaussian Copula

Gaussian copulas are developed in TRD. Other copula distributions are possible, e.g., Student-t distributions (often touted as being more sensitive to fat-tailed distributions — here data is first adaptively fit to fat-tailed distributions prior to copula transformations). These alternative distributions can be quite slow because inverse transformations typically are not as quick as for the present distribution.

Copulas are cited as an important component of risk management not yet widely used by risk management practitioners [4]. Gaussian copulas are presently regarded as the Basel II standard for credit risk management [10]. TRD permits fast as well as robust copula risk management in real time.

The copula approach can be extended to more general distributions than those considered here [11]. If there are not analytic or relatively standard math functions for the transformations (and/or inverse transformations described) here, then these transformations must be performed explicitly numerically in code such as TRD. Then, the ASA_PARALLEL OPTIONS already existing in ASA (developed as part of the1994 National Science Foundation Parallelizing ASA and PATHINT Project (PAPP)) would be very useful to speed up real time calculations [29].

### 4.4.3. Exponential Marginal Distribution Models

For specificity, assume that each market is fit well to a two-tailed exponential density distribution $p$ (not to be confused with the indexed price variable $p_t$) with scale $\chi$ and mean $m$,

$$p(dx)dx = \begin{cases} \dfrac{1}{2\chi} e^{-\frac{dx-m}{\chi}} dx, & dx >= m \\ \dfrac{1}{2\chi} e^{\frac{dx-m}{\chi}} dx, & dx < m \end{cases} = \dfrac{1}{2\chi} e^{-\frac{|dx-m|}{\chi}} dx \qquad (3)$$

which has a cumulative probability distribution

$$F(dx) = \int_{-\infty}^{dx} dx' p(dx') = \dfrac{1}{2}\left[1 + \text{sgn}(dx-m)\left(1 - e^{-\frac{|dx-m|}{\chi}}\right)\right] \qquad (4)$$

where $\chi$ and $m$ are defined by averages $<.>$ over a window of data,

$$m = <dx>, \; 2\chi^2 = <(dx)^2> - <dx>^2 \qquad (5)$$

The exponential distribution is selected here to illustrate that even this hardest case to process analytically [64] can be treated within TRD,

The $p(dx)$ are "marginal" distributions observed in the market, modeled to fit the above algebraic form. Note that the exponential distribution has an infinite number of non-zero cumulants, so that $<dx^2> - <dx>^2$ does not have the same "variance" meaning for this "width" as it does for a Gaussian distribution which has just two independent cumulants (and all cumulants greater than the second vanish). Below algorithms are specified to address correlated markets giving rise to the stochastic behavior of these markets.

The TRD code can be easily modified to utilize distributions $p'(dx)$ with different widths, e.g., different $\chi'$ for $dx$ less than and greater than $m$,

$$p'(dx)dx = \dfrac{1}{2\chi'} e^{-\frac{|dx-m|}{\chi'}} dx \qquad (6)$$

### 4.4.4. Copula Transformation

#### 4.4.4.1. Transformation to Gaussian Marginal Distributions

A Normal Gaussian distribution has the form

$$p(dy) = \dfrac{1}{\sqrt{2\pi}} e^{-\frac{dy^2}{2}} \qquad (7)$$

with a cumulative distribution

$$F(dy) = \dfrac{1}{2}\left[1 + \text{erf}\left(\dfrac{dy}{\sqrt{2}}\right)\right] \qquad (8)$$

where the erf() function is a tabulated function coded into most math libraries.

By setting the numerical values of the above two cumulative distributions, monotonic on interval [0,1], equal to each other, the transformation of the $x$ marginal variables to the $y$ marginal variables is effected,

$$dy = \sqrt{2}\,\text{erf}^{-1}(2\,F(dx)-1) = \sqrt{2}\,\text{sgn}(dx-m)\,\text{erf}^{-1}\left(1 - e^{-\frac{|dx-m|}{\chi}}\right) \tag{9}$$

The inverse mapping is used when applying this to the portfolio distribution,

$$dx = m - \text{sgn}(dy)\chi \ln\left[1 - \text{erf}\left(\frac{|dy|}{\sqrt{2}}\right)\right] \tag{10}$$

### 4.4.4.2. Including Correlations

To understand how correlations enter, look at the stochastic process defined by the $dy^i$ marginal transformed variables:

$$dy^i = \hat{g}^i dw_i \tag{11}$$

where $dw_i$ is the Wiener Gaussian noise contributing to $dy^i$ of market $i$. The transformations are chosen such that $\hat{g}^i = 1$.

Now, a given market's noise, $(\hat{g}^i dw_i)$, has potential contributions from all $N$ markets, which is modeled in terms of $N$ independent Gaussian processes, $dz_k$,

$$\hat{g}^i dw_i = \sum_k \hat{g}^i_k dz_k \tag{12}$$

The covariance matrix $(g^{ij})$ of these $y$ variables is then given by

$$g^{ij} = \sum_k \hat{g}^i_k \hat{g}^j_k \tag{13}$$

with inverse matrix, the "metric," written as $(g_{ij})$ and determinant of $(g^{ij})$ written as $g$.

Since Gaussian variables are now being used, the covariance matrix is calculated directly from the transformed data using standard statistics, the point of this "copula" transformation [64,66].

Correlations $\rho^{ij}$ are derived from bilinear combinations of market volatilities

$$\rho^{ij} = \frac{g^{ij}}{\sqrt{g^{ii} g^{jj}}} \tag{14}$$

Since the transformation to Gaussian space has defined $g^{ii} = 1$, here the covariance matrices theoretically are identical to the correlation matrices.

This gives a multivariate correlated process $P$ in the $dy$ variables, in terms of Lagrangians $L$ and Actions $A$,

$$P(dy) \equiv P(dy^1, \ldots, dy^N) = (2\pi dt)^{-\frac{N}{2}} g^{-\frac{1}{2}} e^{-Ldt} \tag{15}$$

where $dt = 1$ above. The Lagrangian L is given by

$$L = \frac{1}{2dt^2} \sum_{ij} dy^i g_{ij} dy^j \tag{16}$$

The effective action $A_{\text{eff}}$, presenting a "cost function" useful for sampling and optimization, is defined by

$$P(dy) = e^{-A_{\text{eff}}}\ , \quad A_{\text{eff}} = Ldt + \frac{1}{2}\ln g + \frac{N}{2}\ln(2\pi dt) \tag{17}$$

### 4.4.4.2.1. Stable Covariance Matrices

Covariance matrices, and their inverses (metrics), are known to be quite noisy, so often they must be further developed/filtered for proper risk management. The root cause of this noise is recognized as "volatility of volatility" present in market dynamics [60]. In addition to such problems, ill-conditioned matrices can arise from loss of precision for large variables sets, e.g., when calculating inverse matrices and determinants as required here. In general, the window size used for covariance calculations should exceed the number of market variables to help tame such problems.

A very good approach for avoiding ill-conditioning and lack of positive-definite matrices is to perform pre-averaging of input data using a window of three epochs [63]. Other methods in the literature include subtracting eigenvalues of parameterized random matrices [61]. Using Gaussian transformed data alleviates problems usually encountered with fat-tailed distributions. Selection of reasonable windows, coupled with pre-averaging, seems to robustly avoid ill-conditioning.

#### 4.4.4.3. Copula of Multivariate Correlated Distribution

The multivariate distribution in $x$-space is specified, including correlations, using

$$P(dx) = P(dy) \left| \frac{\partial\, dy^i}{\partial\, dx^j} \right| \tag{18}$$

where $\left| \frac{\partial dy^i}{\partial dx^j} \right|$ is the Jacobian matrix specifying this transformation. This gives

$$P(dx) = g^{-\frac{1}{2}} e^{-\frac{1}{2} \sum_{ij} (dy^i_{dx})^\dagger (g_{ij} - I_{ij})(dy^j_{dx})} \prod_i P_i(dx^i) \tag{19}$$

where $(dy_{dx})$ is the column-vector of $(dy^1_{dx}, \cdots, dy^N_{dx})$ expressed back in terms of their respective $(dx^1, \ldots, dx^N)$, $(dy_{dx})^\dagger$ is the transpose row-vector, and $(I)$ is the identity matrix (all ones on the diagonal). The Gaussian copula $C(dx)$ is defined from Eq. (19),

$$C(dx) = g^{-\frac{1}{2}} e^{-\frac{1}{2} \sum_{ij} (dy^i_{dx})^\dagger (g_{ij} - I_{ij})(dy^j_{dx})} \tag{20}$$

### 4.4.5. Portfolio Distribution

The probability density $P(dM)$ of portfolio returns $dM$ is given as

$$P(dM) = \int \prod_i d(dx^i) P(dx) \delta_D(dM_t - \sum_j (a_{j,t}\, dx^j + b_{j,t})) \tag{21}$$

where the Dirac delta-function $\delta_D$ expresses the constraint that

$$dM = \sum_j (a_j\, dx^j + b_j) \tag{22}$$

The coefficients $a_j$ and $b_j$ are determined by specification of the portfolio current $K_{t'}$, and "forecasted" $K_t$, giving the returns expected at $t$, $dM_t$,

$$dM_t = \frac{K_t - K_{t'}}{K_{t'}}$$

$$K_{t'} = Y_{t'} + \sum_i \text{sgn}(NC_{i,t'}) NC_{i,t'}(p_{i,t'} - p_{i,@,t'})$$

$$K_t = Y_t + \sum_i (\text{sgn}(NC_{i,t}) NC_{i,t}(p_{i,t} - p_{i,@,t}) + SL[NC_{i,t} - NC_{i,t'}]) \tag{23}$$

where $NC_{i,t}$ is the current number of broker-filled contracts of market $i$ at time $t$ ($NC > 0$ for long and $NC < 0$ for short positions), $p_{i,@,t'}$ and $p_{i,@,t}$ are the long/short prices at which contracts were bought/sold according to the long/short signal $\text{sgn}(NC_{i,t'})$ generated by external models. $Y_t$ and $Y_{t'}$ are the dollars available for investment. The function $SL$ is the slippage and commissions suffered by changing the number of contracts.

#### 4.4.5.1. Recursive Risk-Management in Trading Systems

Sensible development of trading systems fit trading-rule parameters to generate the "best" portfolio (best depends on the chosen criteria). This necessitates fitting risk-managed contract sizes to chosen risk targets, for each set of chosen trading-rule parameters, e.g., selected by an optimization algorithm. A given set of trading-rule parameters affects the $a_{j,t}$ and $b_{j,t}$ coefficients in Eq. (21) as these rules act on the forecasted market prices as they are generated to sample the multivariate market distributions.

This process must be repeated as the trading-rule parameter space is sampled to fit the trading cost function, e.g., based on profit, Sharpe ratio, etc., of the Portfolio returns over a reasonably large in-sample set of data.

### 4.4.6. Risk Management

Once $P(dM)$ is developed (e.g., numerically), risk-management optimization is defined. The portfolio integral constraint is,

$$Q = P(dM < VaR) = \int_{-\infty}^{-|VaR|} dM \, P(M_t | M'_{t'}) \tag{24}$$

where $VaR$ is a fixed percentage of the total available money to invest. E.g., this is specifically implemented as

$$VaR = 0.05 \, , \, Q = 0.01 \tag{25}$$

where the value of $VaR$ is understood to represent a possible 5% loss in portfolio returns in one epoch, e.g., which approximately translates into a 1% chance of a 20% loss within 20 epochs. Expected tail loss (ETL), sometimes called conditional VaR or worst conditional expectation, can be directly calculated as an average over the tail. While the VaR is useful to determine expected loss if a tail event does not occur, ETL is useful to determine what can be lost if a tail event occurs [7].

ASA [29] is used to sample future contracts defined by a cost function, e.g., maximum profit, subject to the constraint

$$Cost_Q = |Q - 0.01| \tag{26}$$

by optimizing the $NC_{i,t}$ parameters. Other post-sampling constraints can then be applied. (Judgments always must be made whether to apply specific constraints, before, during or after sampling.)

Risk management is developed by (ASA-)sampling the space of the next epoch's $\{NC_{i,t}\}$ to fit the above $Q$ constraint using the sampled market variables $\{dx\}$. The combinatoric space of $NC$'s satisfying the $Q$ constraint is huge, and so additional $NC$-models are used to choose the actual traded $\{NC_{i,t}\}$.

### 4.4.7. Sampling Multivariate Normal Distribution for Events

Eq. (21) certainly is the core equation, the basic foundation, of most work in risk management of portfolios. For general probabilities not Gaussian, and when including correlations, this equation cannot be solved analytically.

Some people approximate/mutilate this multiple integral to attempt to get some analytic expression. Their results may in some cases serve as interesting "toy" models to study some extreme cases of variables, but there is no reasonable way to estimate how much of the core calculation has been destroyed in this process.

Many people resort to Monte Carlo sampling of this multiple integral. ASA has an ASA_SAMPLE option that similarly could be applied. However, there are published algorithms specifically for multivariate Normal distributions [8].

#### 4.4.7.1. Transformation to Independent Variables

The multivariate correlated $dy$ variables are further transformed into independent uncorrelated Gaussian $dz$ variables. Multiple Normal random numbers are generated for each $dz^i$ variable, subsequently transforming back to $dy$, $dx$, and $dp$ variables to enforce the Dirac $\delta$-function constraint specifying the $VaR$ constraint.

The method of Cholesky decomposition is used (eigenvalue decomposition also could be used, requiring inverses of matrices, which are used elsewhere in this project), wherein the covariance matrix is factored into a product of triangular matrices, simply related to each other by the adjoint operation. This is possible because $G$ is a symmetric positive-definite matrix, i.e, because care has been taken to process the raw data to preserve this structure as discussed previously.

$$G = (g^{ij}) = C^{\dagger} C \ , \ I = C \, G^{-1} \, C^{\dagger} \tag{27}$$

from which the transformation of the $dy$ to $dz$ are obtained. Each $dz$ has 0 mean and StdDev 1, so its covariance matrix is 1:

$$I = <(dz)^{\dagger}(dz)> = <(dz)^{\dagger} (C \, G^{-1} \, C^{\dagger})(dz)> = <(C^{\dagger} dz)^{\dagger} \, G^{-1} \, (C^{\dagger} dz)> = <(dy)^{\dagger} \, G^{-1} \, (dy)> \tag{28}$$

where

$$dy = C^{\dagger} \, dz \tag{29}$$

The collection of related $\{dx\}$, $\{dy\}$, and $\{dz\}$ sampled points are defined here as Events related to market movements.

### 4.4.8. Numerical Development of Portfolio Returns

#### 4.4.8.1. X From Sampled Events Into Bins

One approach is to directly develop the portfolio-returns distribution, from which moments are calculated to define $Q$. This approach has the virtue of explicitly exhibiting the shapes of the portfolio distribution being used. In some production runs, integration over the Dirac $\delta$-function permits faster numerical calculations of moments of the portfolio distribution, to fit these shapes.

The sampling process of Events are used to generate portfolio-return Bins to determine the shape of $P(dM)$. Based on prior analyses of data — market distributions have been assumed to be basically two-tailed exponentials — here too prior analyses strongly supports two-tailed distributions for the portfolio returns. Therefore, only a "reasonable" sampling of points of the portfolio distribution, expressed as Bins, is needed to calculate the moments. For example, a base function to be fitted to the Bins would be in terms of parameters, width X and mean $m_M$,

$$P(dM)dM = \begin{cases} \dfrac{1}{2X} e^{-\frac{dM-m_M}{X}} dM \ , \ dM >= m_M \\ \dfrac{1}{2X} e^{\frac{dM-m_M}{X}} dM \ , \ dM < m_M \end{cases} = \dfrac{1}{2X} e^{-\frac{|dM-m_M|}{X}} dM \tag{30}$$

X and $m_M$ are defined from data in the Bins by

$$m_M = <dM> \ , \ 2X^2 = <(dM)^2> - <dM>^2 \tag{31}$$

By virtue of the sampling construction of $P(dM)$, X implicitly contains all correlation information inherent in $A'_{eff}$.

The TRD code can be easily modified to utilize distributions $P'(dM)$ with different widths, e.g., different $X'$ for $dM$ less than and greater than $m_M$,

$$P'(dM)dM = \dfrac{1}{2X'} e^{-\frac{|dM-m_M|}{X'}} dM \tag{32}$$

A large number of Events populate Bins into the tails of $P(dM)$. Different regions of $P(dM)$ could be used to calculate a piecewise X to compare to one X over the full region, with respect to sensitivities of values obtained for $Q$,

$$Q = \dfrac{1}{2} e^{-\frac{|VaR-m_M|}{X}} \tag{33}$$

Note that fixing $Q$, $VaR$, and $m_M$ fixes the full shape of the portfolio exponential distribution. Sampling of the $NC_i$ is used to adapt to this shape constraint.

### 4.4.9. Multiple Trading Systems

TRD is designed to process multiple trading systems. A top-level text parameter file read in by the running code adaptively decides which trading systems to include at any upcoming epoch, without requiring recompilation of code.

For example, a master controller of system libraries could change this parameter file at any time so that at the next epoch of real-time trading a new set of systems could be in force, or depending on the markets contexts a set of top-level master-controller parameters could decide in training (and used for real-time this way as well) which libraries to use. The flag to include a system is a number which serves as the weight to be used in averaging over signals generated by the systems prior to taking a true position. This approach permits the possibility of encasing all trading systems in a global risk-management and a global optimization of all relevant trading-rule parameters.

TRD is designed to easily insert and run multiple trading systems, e.g., to add further diversification to risk-managing a portfolio. Some trading systems may share indicators and parameters, etc.

## 5. Future Research

### 5.1. Anticipated Results

If this project is successful it will have been established that SMNI and TRD algorithms, supported by computational algorithms such as ASA and PATHINT/PATHTREE, is a very viable approach to develop Ideas by Statistical Mechanics (ISM) for decision support for DIME factors of propagation/evolution of ideas.

### 5.2. Significance of Initial Efforts

Endogenous and exogenous interactions among local populations, one or some of which have been fit to an Idea(s) will be tuned to determine circumstances under which the Idea(s) can be propagated or stopped. It is possible that only populations already approximately fit/prepared for the Ideas(s) may be most receptive, and the degree of such preparedness must be determined. If SMNI is any guide, it is easy to see how long-ranged connectivity can sometimes be an effective substitute for having tuned local interactions.

## 6. Commercialization

This product can be used for decision support for projects ranging from diplomatic, information, military, and economic (DIME) factors of propagation/evolution of ideas, to commercial sales, trading indicators across sectors of financial markets, advertising and political campaigns, etc.

Lester Ingber Research                     - 21 -                                    ISM